\documentclass[12pt]{iopart}
\usepackage{iopams} 

\newcommand{\supp}{ {\rm supp} }
\newcommand{\mean}[1]{\langle #1\rangle}

\newcommand{\R}{{\mathbb R}}

\newcommand{\order}[1]{{\mathcal O}(#1)}
\newcommand{\vep}{\varepsilon}
\newcommand{\vc}[1]{{\bf #1}}

\newcommand{\defset}[2]{ \left\{ #1\left|\, #2 %
  \makebox[0cm]{$\displaystyle\phantom{#1}$}\right.\!\right\} }
\newcommand{\norm}[1]{\Vert #1\Vert}
\newcommand{\re}{{\rm Re\,}}

\newcommand{\ope}[1]{\widehat{#1}}
\newcommand{\half}{{1\over 2}}
\newcommand{\tihop}{\ope{\rho}}
\newcommand{\ham}{\ope{H}}

\newcommand{\tihep}{\tihop_\varepsilon}

\newcommand{\gmean}[1]{\mean{#1}^{{\rm gauss}}}
\newcommand{\canmean}[1]{\mean{#1}^{{\rm can}}}
\newcommand{\mcmean}[1]{\mean{#1}^{{\rm micro}}}
\newcommand{\ZG}{Z^{{\rm gauss}}}
\newcommand{\Zcan}{Z^{{\rm can}}}
\newcommand{\Zmc}{Z^{{\rm micro}}}
\newcommand{\SG}{S^{{\rm gauss}}}
\newcommand{\Scan}{S^{{\rm can}}}

\newcommand{\ase}{{\sigma^2\over 2\vep^2}}
\newcommand{\text}[1]{\makebox[\width]{#1}}
\newcommand{\overset}[2]{\stackrel{#1}{#2}}

\begin{document} 


\jl{1}

\title[Microcanonical improvement of the quantum canonical ensemble]{%
Derivation and improvements of the quantum canonical ensemble from a 
regularized microcanonical ensemble}
\author{Jani Lukkarinen\footnote[1]{Electronic mail:
 {\tt jani.lukkarinen@helsinki.fi} }}
\address{Helsinki Institute of Physics, P.O.Box 9, 
00014 University of Helsinki, Finland} 

\begin{abstract}
We develop a regularization of the quantum microcanonical ensemble,
called a Gaussian ensemble, which can be used for derivation of the
canonical ensemble from microcanonical principles.  The derivation
differs from the usual methods by giving an explanation for the, at the
first sight unreasonable, effectiveness of the canonical ensemble when
applied to certain small, isolated, systems.  This method also allows a
direct identification between the parameters of the microcanonical and
the canonical ensemble and it yields simple indicators and rigorous
bounds for the effectiveness of the approximation.  Finally, we derive
an asymptotic expansion of the microcanonical corrections to the
canonical ensemble for those systems, which are near, but not quite, at
the thermodynamical limit and show how and why the canonical ensemble
can be applied also for systems with exponentially increasing density of
states.  The aim throughout the paper is to keep mathematical rigour
intact while attempting to produce results both physically and
practically interesting. 
\end{abstract}


\submitted
\maketitle

\section{Introduction}

The microcanonical ensemble is considered to be the fundamental ensemble
of statistical physics.  For example, the use of the canonical Gibbs
ensemble is usually justified by showing that its expectation values
coincide with the microcanonical ones in the thermodynamical limit,
when the size of the system approaches infinity.  However, since an
application of the microcanonical ensemble requires detailed knowledge
about the energy levels of the system, it is seldom possible to use it
in practice.  The canonical ensemble, on the other hand, has a
well-behaving path-integral expression easily extended to gauge field
theories---therefore it has become the standard ensemble of quantum
statistics. 

Nevertheless, there are situations where the easiest, canonical,
alternative does not work.  For instance, direct applications of the
grand canonical ensemble have not been able to reproduce all the results
of relativistic ion collision experiments \cite{redlich:94}. There
are two possible explanations for this shortcoming of the canonical
ensemble: either the particle gas created in the collision does not
reach thermal equilibrium before exploding into the final state hadrons,
or the system is too small to be handled by canonical methods. 
In fact, recent calculations \cite{braunm:96} using 
a modified grand canonical ensemble have succeeded in describing
most of the particle abundances in these experiments, but only at the
cost of including finite-volume corrections to the usual ensemble.  
This suggests
that at least the final state hadron gas will thermalize, but it also
shows that finite-volume effects are prominent in these systems. 

These results point into the direction that the microcanonical
ensemble---which is optimal for describing isolated ergodic systems with
small quantum numbers---should be used for getting quantitative
information about the properties of the quark-gluon plasma possibly
created in relativistic hadron collisions.  There are already proposals
how this can be accomplished in the continuum path-integral formulation
of field theories \cite{chs:93}, but since the argumentation in
this kind of formalism cannot be made completely mathematically rigorous,
a further study of the details of the 
quantum microcanonical ensemble was felt to be necessary. 

When trying to do rigorous quantum microcanonical computations, one
immediately encounters two practical difficulties associated with the
discrete spectrum of an isolated quantum system: since the spectrum is
discrete, the possible values of the spectrum have to be known in
advance before any computations can be done; on the other hand, the
position of the high energy spectral levels depend on small fluctuations
of the interaction potential.  Consider, for example, the harmonic
oscillator, for which the energy levels are $E_n = \omega (n + \half)$:
for $n=100$ one percent change in the oscillator frequency $\omega$ will
change the position of the level $E_{100}$ by a whole energy unit!
Therefore, the spectral levels by themselves are not very practical
parameters; however, the {\em density}\/ of the spectral levels is
robust in fluctuations of the potential and thus a smoothened energy
spectrum would offer a more stable alternative. 

This paper attempts to fill the gap between the canonical and the
microcanonical ensemble in quantum mechanics.  The main ingredient in
this is the introduction of the Gaussian ensemble, which is essentially
just a regularization of the discrete spectrum of a closed quantum
system.  The Gaussian ensemble will be defined in section two, where we
will also show how all microcanonical results can be obtained from the
Gaussian ensemble in the limit where the regularization is removed.  In
the following sections we will then show that in a certain range of the
regularization parameter the canonical ensemble is a good approximation
of the Gaussian ensemble and we will propose definite ways of estimating
the difference between the two methods.  These results are then employed
in the derivation of an asymptotic expansion of microcanonical
corrections to the canonical ensemble, which will be most useful for
systems near, but not quite at, the thermodynamical limit.  Finally, we
will give some implications of the present results to the thermodynamics
of systems with exponentially increasing density of states, for which
the canonical ensemble is in principle not well-defined. 

We will not repeat standard results or definitions of statistics of
quantum systems in the thermodynamical limit here.  The physical
argumentation leading to the density operator formulation is explained
in most textbooks on the subject, volume five of the classical series by
Landau and Lifshitz \cite{ll:statphys} being a good example.  More
elaborate and recent analysis of the subject can be found from volume
one of the series by Balian \cite{balian1}, while a mathematically
rigorous approach is developed in volume four of ``A Course in
Mathematical Physics'' by Thirring \cite{thirring4}. 


\section{Gaussian ensemble\protect\footnote{
On physical grounds, the idea of using a Gaussian energy distribution to
define a statistical ensemble seems natural.  Such a physical reasoning
was adopted, for instance, in \cite{HETH:87} to introduce
a Gaussian ensemble for studying first order 
phase transitions in certain lattice gauge theories. }
as a regularization of the microcanonical ensemble}

Consider the operator defined by
\begin{equation} \label{e:tihep} 
\tihep(E) = {1\over \sqrt{2 \pi
\varepsilon^2} } \exp\!\!\left[-\half \left( {\ham -E \over \varepsilon}
\right)^2 \right], 
\end{equation} 
where $\ham$ is the Hamiltonian and $E$ and $\vep\ne 0$ are two real
parameters.  Since $\ham$ is self-adjoint, $\tihep(E)$ is bounded,
self-adjoint and positive.  For this operator to define a sensible
statistical ensemble via the trace formulas, it is also necessary to
require that the spectrum of the Hamiltonian is discrete and increases
sufficiently fast at infinity so that $\Tr \tihep(E) < \infty$.  If this
is true, we will define the {\em Gaussian expectation values}\/ of an
observable $\ope{A}$ by the formula
\begin{equation}\label{e:defev} 
\gmean{\ope{A}}_{E,\vep} \equiv 
 {\Tr\!\left(\ope{A} \tihep(E)\right) \over \Tr\tihep(E)}.
\end{equation} 

At this point, it will be useful to define some terminology to be used
later.  If the Hamiltonian of the system satisfies $\Tr \tihep(E) <
\infty$ for all $E\in\R$ and $\vep>0$, we will call the system {\em
Gaussianly bounded}\/ and define the above trace as the Gaussian
partition function $\ZG_{E_,\vep}$.  Similarly, if $\Tr \rme^{-\beta
\ham}<\infty$ for all $\beta >0$, we will say that the system is {\em
canonically bounded}\/ and the trace will give the canonical partition
function $\Zcan_\beta$.  The terms {\em Gaussian}\/ and {\em canonical
observable,}\/ respectively, are then used for those normal or bounded
operators (i.e.\ observables) $\ope{A}$ for which
$\gmean{|\ope{A}|}_{E_,\vep} < \infty$ or 
$\canmean{|\ope{A}|}_\beta<\infty$ 
for all the above values of the parameters---here $|\ope{A}|$
refers to the positive square root of $\ope{A}^\dagger\ope{A}$. 
Analogously, systems with a discrete energy spectrum with finite
multiplicities are called microcanonically bounded and all normal
operators having a domain which contains the domain of the Hamiltonian
as well as all bounded operators are microcanonical observables.  In
this last case, the microcanonical partition function $\Zmc_E$ is
defined by the number of eigenstates with energy $E$ and the
microcanonical expectation values are the averages over these
eigenstates. 

Clearly, every canonically bounded system is Gaussianly bounded and every
canonical observable is a Gaussian observable, and the same relations hold
between the Gaussian and the microcanonical ensemble as well. The
term bounded used here has also 
a direct physical interpretation in the canonical case:
by the Golden-Thompson-Symanzik inequality \cite{GTS}, which 
applies for $n$-dimensional systems with Hamiltonians of the form 
$\ham= \half \ope{\vc{p}}^2 + V(\ope{\vc{x}})$, a system is canonically
bounded if the potential increases 
fast enough at infinity so that 
$\int\! \rmd^n\vc{x}\, \rme^{-\beta V(\vc{x})} <\infty$ 
for all $\beta >0$.

The formal relation $\lim_{\varepsilon \to 0} \tihep(E) 
= \delta(\ham - E)$ 
is one obvious motivation for using this ensemble to approximate the
microcanonical one.  However, it can also be argued that the Gaussian
ensemble is even better suited for describing typical experimental
situations than the microcanonical ensemble: if a quantum system was
initially prepared into, or was measured to have, an energy $E$ and the
system is almost, but not completely, isolated having interactions with
the environment that lead to energy fluctuations of the order of $\vep$,
then the Gaussian ensemble using these parameters is the most natural
way to predict the behaviour of a statistical average over many
independent measurements of an observable.  Of course, for this physical
interpretation to be valid, the interactions with the environment need
to be balanced in such a way as not to lead to a net flow of energy from
one direction to the other---this is the essence of the requirement of
``thermalization'' of the system in the context of the Gaussian
ensemble. 

Mathematically, the Gaussian ensemble is a regularization of the
discrete energy spectrum by a convolution with the normal
distribution.  As was explained in the introduction, this is beneficial
since it offers a way of removing the effect of the instability of the high
energy spectral levels. Also, its use does not require any prior knowledge
about the spectrum as it is well defined for all values of $E$.
Most important is, however, the way how the 
Gaussian ensemble offers a natural and mathematically rigorous
approximation of the microcanonical ensemble:
\begin{enumerate}
\item $\lim_{\varepsilon \to 0}\sqrt{2 \pi \vep^2} \ZG_{E,\vep} 
	= \Zmc_E$ for all real $E$.
\item $\lim_{\varepsilon \to 0} \ZG_{E,\vep} = \Tr \delta(\ham-E)$ as
 distributions in $E$.
\item $\lim_{\varepsilon \to 0} \gmean{\ope{A}}_{E,\vep} =
	\mcmean{\ope{A}}_{E'}$, where $E'$ is the energy eigenvalue
	nearest to $E$.
\end{enumerate}
We will now conclude this section by proving these three statements.

Suppose that the system is Gaussianly bounded with discrete
energy levels $E_n$, each one having a multiplicity $\kappa_n$ and
eigenvectors $\Omega_{n,k}$, $k=1,\ldots,\kappa_n$.  
Then for any Gaussian observable $\ope{A}$,
the trace in (\ref{e:defev}) can be expressed as
\begin{equation}
\label{e:tracesum}
\Tr\!\left(\ope{A} \tihep(E)\right) = 
 {1\over \sqrt{2 \pi \varepsilon^2} }
 \sum_{n,k} \exp\!\left[ -{1\over 2 \vep^2}
  \left( E_n -E \right)^2 \right] 
  \langle\Omega_{n,k}|\widehat{A}|\Omega_{n,k}\rangle. 
\end{equation}
Denote $\langle\Omega_{n,k}|\widehat{A}|\Omega_{n,k}\rangle$ by
$a_{n,k}$ and note that since $|a_{n,k}|\le
\langle\Omega_{n,k}|\,|\widehat{A}|\,|\Omega_{n,k}\rangle$, the above
series converges absolutely for any $\vep>0$.  

Let us first look at the behaviour of the Gaussian partition function,
i.e.\ use $\ope{A}=\ope{1}$.  By (\ref{e:tracesum}) then
\[
\sqrt{2 \pi \vep^2} \ZG_{E,\vep} = \sum_n \kappa_n 
  \exp\!\left[ -{1\over 2 \vep^2} \left( E_n -E \right)^2 \right].
\]
Since for any $W>0$ the function $\rme^{-W/\vep^2}$ is an increasing
function of $\vep$ in the region $\vep>0$, dominated convergence can be
invoked to move the limit $\vep\to 0$ inside the sum, which then gives
the result
\[
\lim_{\varepsilon \to 0}\sqrt{2 \pi \vep^2} \ZG_{E,\vep} = 
\left\{\begin{array}{l}
   \kappa_n\mbox{, if }E=E_n\text{ for some }n \\
   0\text{, otherwise} 
 \end{array}\right..
\]
Since the right hand side equals $\Zmc_E$ by definition, this proves
the first statement.
	
Let next $f(E)$ be any smooth function with a compact support and let
$M>0$ be such that $|\supp f|\le M$.  To prove the second
statement, we need to show that
\[
\lim_{\vep \to 0} \int_{-\infty}^\infty \rmd E\, f(E) \ZG_{E,\vep}
 = \sum_n \kappa_n f(E_n).
\]
Since $\int_{-\infty}^\infty \!\rmd E\, f(E) {1\over \sqrt{2 \pi\vep^2}}
  \exp\!\left[ -{1\over 2 \vep^2}
  \left( E_n -E \right)^2\right] \to f(E_n)$, when $\vep \to 0$, this
is clearly true if only it were possible to move first the
$E$-integration and then the $\vep\to 0$ limit inside the sum.  But, in
fact, both of these operations are now allowed by the dominated
convergence theorem, since we have the bounds
\begin{eqnarray}
\fl
\int_{-\infty}^\infty \!\rmd E\, |f(E)| 
 { \exp\!\left[ -{1\over 2 \vep^2}
   \left( E_n -E \right)^2\right] \over \sqrt{2 \pi\vep^2} } \nonumber\\
 \lo{\le} \left\{\begin{array}{l}
   \norm{f}_\infty\text{, if }|E_n|<2 M \\
   \norm{f}_\infty \exp\!\left[ -\half \left( |E_n|-M \right)^2 + \half
 	M^2\right]\text{, if }|E_n|\ge 2 M  
 \end{array}\right., \nonumber
\end{eqnarray}
for all $0<\vep \le 1$. 

Let us finally evaluate the limit of the expectation values for general
$\ope{A}$,
\begin{equation}
\label{e:meanlim}
\gmean{\ope{A}}_{E,\vep} = \sum_{n,k}
 {\exp\!\left[{ -{1\over 2\vep^2} \left( {E_{n} -E } \right)^2 }\right]
 \over  \Tr \tihep \,\sqrt{2 \pi \varepsilon^2} } 
 \langle\Omega_{n,k}| \widehat{A} |\Omega_{n,k}\rangle.
\end{equation}
For this we will need the result
\begin{equation}\label{e:elim}
\fl
{ 1 \over \sum\limits_{n',k'} \exp\!\left[ {
     -{(E_{n'} -E)^2 - (E_n -E)^2 \over 2 \varepsilon^2} } \right] } 
 \overset{\vep\to 0}{\longrightarrow} \left\{\begin{array}{l}
   0\mbox{, if }|E_{n'} -E|< |E_n -E|\mbox{ for some }n' \\
   (\sum_{n'} \kappa_{n'}\delta_{|E_{n'} -E|,|E_n -E|})^{-1}
     \mbox{, otherwise} 
 \end{array}\right. .
\end{equation}

In other words, this limit is zero if $E_n$ is not the eigenvalue
nearest to $E$, it is $(\kappa_n+\kappa_{m})^{-1}$ if both $E_n$ and
$E_{m}$ are nearest eigenvalues (i.e.\ if $E$ lies exactly in the middle
of the segment joining $E_n$ and $E_{m}$ and no other eigenvalues are on
this segment) and it is $\kappa_n^{-1}$ if $E_n$ is a unique nearest
eigenvalue.  Let us use notation $M_n(E)$ for the sum 
$\sum_{n'} \kappa_{n'}\delta_{|E_{n'} -E|,|E_n -E|}$. 

Let $n_0$ be the index of any one of the eigenvalues nearest to $E$.
Then for all $n$
\begin{equation}
\label{e:exineq}
\fl
 {\exp\!\left[{ -{1\over 2\vep^2} \left( {E_{n} -E } \right)^2 }\right]
  \over \Tr \tihep \,\sqrt{2 \pi \varepsilon^2} } =
{ \exp\!\left[{ -{1\over 2\vep^2} \left( {E_{n} -E } \right)^2 }\right]
  \over \sum\limits_{n'} \kappa_{n'} \exp\!\left[ {
     -{(E_{n'} -E)^2 \over 2 \varepsilon^2} }\right] } 
< {1\over \kappa_{n_0}} \exp\!\left[ { (E_{n_0} -E)^2 - (E_n -E)^2
	\over 2 \vep^2} \right],
\end{equation}  
where $(E_{n_0} -E)^2 - (E_n -E)^2 \le 0$.  This implies that,
for all $\vep$ in the range $0<\vep\le 1$, the absolute value of 
each of the terms of the series in (\ref{e:meanlim}) is less than 
$\exp[{(E_{n_0} -E)^2 / 2}]
 \exp[{-(E_n -E)^2 / 2 }] |a_{n,k}|$, which again form an
$\varepsilon$-independent sum that is convergent by assumption.  
Thus dominated convergence can be applied to move the limit inside the
sum, which then by equation (\ref{e:elim}) yields the result
\[
\gmean{\ope{A}}_{E,\vep} 
 \overset{\vep\to 0}{\longrightarrow} {1\over M_{n_0}(E)}
 \sum_{n,k} \delta_{|E_{n} -E|,|E_{n_0} -E|}
  \langle\Omega_{n,k}| \widehat{A} |\Omega_{n,k}\rangle.
\]
Since $M_{n_0}(E)$ is the number of non-zero
terms in the above sum, the final expression
is nothing but the average of the expectation values of $\widehat{A}$ over
the energy eigenstates nearest to $E$.

Therefore, whenever $E$ coincides with a point in the spectrum, the
limit $\varepsilon\to 0$ will give the microcanonical expectation value. 
On the other hand, if $E$ does not belong to the energy spectrum (in
which case the microcanonical ensemble is in principle ill-defined),
then the microcanonical result corresponding to the nearest eigenvalue
is obtained.  The only values of $E$ giving non-microcanonical limits
are those lying exactly in the middle between two eigenvalues, but even
then the result is an expectation value of a uniform distribution over
two energy eigenvalues. 

\section{Canonical ensemble as an approximation to the Gaussian
ensemble\label{sec:cangauss}} 

We will next show how the canonical ensemble can be used for
approximating the Gaussian ensemble in the region where the energy
resolution $\vep$ is sufficiently large.  For this we need to assume
that the system is canonically bounded and that $\ope{A}$ is a canonical
observable.  Since this, in particular, requires the energy spectrum to
be bounded from below, we will also assume that the Hamiltonian has been
normalized so that the lowest energy level $E_0$ is non-negative. 

Let us first assume that $\beta$ is a positive parameter.  Since
\[
-{1\over 2 \vep^2} (E-E_n)^2 = -{1\over 2 \vep^2} (E-E_n+\beta\vep^2)^2
  + \beta E -\beta E_n + \half \beta^2 \vep^2,
\]
we have the identity
\begin{equation}\label{e:gid}
\Tr\!\left(\ope{A} \tihep(E)\right) 
 = \rme^{\half \beta^2\vep^2+\beta E}
 \Tr\!\left(\widehat{A} \tihep(E+\beta\vep^2) \rme^{-\beta\ham}\right).
\end{equation}
If we integrate this multiplied by $\rme^{-\beta E}$ over $E$ and
take the integration inside the trace, which is possible 
since $\ope{A}$ is a canonical observable, we get the exact formula
\begin{equation}\label{e:gauscan}
\Tr\!\left(\ope{A} \rme^{-\beta\ham}\right) 
= \rme^{-\half \beta^2\vep^2}
\int_{-\infty}^\infty \!\! \rmd E\, \rme^{-\beta E} 
\Tr\!\left(\ope{A}\tihep(E)\right)
\end{equation}
valid for all $\vep>0$ and $\beta>0$.  This is a regularized form of
the familiar statement that the canonical ensemble is the Laplace
transform of the microcanonical one.

The result (\ref{e:gauscan}) has a more interesting inverse
formula, which we will derive next.  We will begin with
the Fourier-transform of the normal distribution,
\begin{equation}
\label{e:gaussint}
\int_{-\infty}^\infty {d\alpha\over 2\pi}
\exp\!\left({-\half \varepsilon^2\alpha^2 + \rmi \alpha W}\right) =
{1\over \sqrt{2 \pi \varepsilon^2} } \exp\!\left({-{1\over 2 \vep^2}
{ W^2 } }\right),
\end{equation}
which can be applied to (\ref{e:gid}), yielding
\[
\Tr\!\left(\ope{A} \tihep(E)\right) 
 = \rme^{\half \beta^2\vep^2+\beta E}
 \int_{-\infty}^\infty {d\alpha\over 2\pi} 
 \rme^{-\half \varepsilon^2\alpha^2} 
 \Tr\!\left( \widehat{A} 
   \rme^{-\beta\ham+\rmi\alpha(E+\beta\vep^2-\ham)}\right).
\]
In changing the order of the integration and the trace we again had to
use the assumption that $\ope{A}$ is a canonical observable.  Using now
a new integration variable $w=\beta+\rmi \alpha$ we get a particularly
simple form of the desired inversion formula for (\ref{e:gauscan}),
\begin{equation}\label{e:invform}
\Tr\!\left(\ope{A} \tihep(E)\right) = 
  \int_{\beta-\rmi\infty}^{\beta+\rmi\infty} {dw\over 2\pi\rmi}
  \rme^{\half\vep^2 w^2 + w E} \Tr\!\left(\ope{A}\rme^{-w\ham}\right)
\end{equation}
valid for all $\beta>0$.  We can now conclude that the analytical form
of the canonical trace contains all the information needed to compute
the microcanonical expectation values, which are then easily extracted
from the integral in (\ref{e:invform}).  However, this formula leads
also to a simple relation between the usual real-temperature canonical
ensemble and the Gaussian ensemble which we shall inspect next. 

For all canonical observables, the trace
$\Tr\!\left(\ope{A}\rme^{-w\ham}\right)$ is obviously an analytic
function of $w$ in the half-plane $\re w >0$ and all its derivatives are
given by a differentiation inside the trace, i.e.\
\[
{\rmd^k\over \rmd w^k} \Tr\!\left(\ope{A}\rme^{-w\ham}\right) =
  \Tr\!\left(\ope{A}(-\ham)^k \rme^{-w\ham}\right).
\]  
Therefore, saddle point methods can be used in evaluation of the
integral in (\ref{e:invform}),
\[
  \int_{\beta-\rmi\infty}^{\beta+\rmi\infty} {dw\over 2\pi\rmi}
  \exp\!\!\left[
   \half\vep^2 w^2 + w E + \ln\Tr\!\left(\ope{A}\rme^{-w\ham}\right)
  \right].
\]
Here the branch of the logarithm needs to be chosen so that the
logarithm is analytic on the integration contour; if the contour happens
to go through a zero of the trace, then an infinitesimal deformation of
the contour is necessary---note that this is always possible if we only
make the trivial assumption $\ope{A}\ne 0$.  The saddle point equation,
which is of course independent of what branch we use for the logarithm,
is
\begin{equation}\label{e:speq}
 { \Tr\!\left(\ope{A}\ham \rme^{-w\ham}\right) \over
      \Tr\!\left(\ope{A}\rme^{-w\ham}\right) }  = E + \vep^2 w,
\end{equation}
while the second derivative, which will determine 
the direction of the the steepest descent path, is
\begin{equation}\label{e:sp2}
 { \Tr\!\left(\ope{A}\ham^2 \rme^{-w\ham}\right) \over
      \Tr\!\left(\ope{A}\rme^{-w\ham}\right) } -
 \Biggl( { \Tr\!\left(\ope{A}\ham \rme^{-w\ham}\right) \over
      \Tr\!\left(\ope{A}\rme^{-w\ham}\right) } \Biggr)^2
 + \vep^2.
\end{equation}

So far we have assumed the observable $\ope{A}$ only to be
a canonical one.  For the following discussion we shall also assume
that $\ope{A}$ is positive, non-zero and that 
$\Tr\!\left(\ope{A}\ham\right)=\infty$, keeping in mind that the special
case $\ope{A}=\ope{1}$ falls into this category.  We shall also
use the notation 
$\overline{O}$ for $\Tr\!\left(\ope{A}\ope{O} \rme^{-w\ham}\right)/
\Tr\!\left(\ope{A}\rme^{-w\ham}\right)$.  With these definitions, the
saddle point equation becomes $\overline{H} = E + \vep^2 w$ and the
second derivative is $\overline{(H-\overline{H})^2} + \vep^2$.

If we restrict ourselves to the positive real axis, then 
$\Tr\!\left(\ope{A}\rme^{-w\ham}\right)$ will always be strictly 
positive and both $\overline{H}$ and $\overline{(H-\overline{H})^2}$
will be well-defined and positive.  From this we conclude that on the
positive real axis the logarithm 
$\ln \Tr\!\left(\ope{A}\rme^{-w\ham}\right)$
is a convex function and that the expectation value $\overline{H}$ is
strictly decreasing.  Therefore, the saddle point equation
(\ref{e:speq}) has at most one positive solution for each $E$.  On the
other hand, the assumption $\Tr\!\left(\ope{A}\ham\right)=\infty$ can
be used for showing that a solution exists for every $E\in\R$ and
$\vep>0$.  We shall now assume that $\beta$, which was an
arbitrary parameter of the integration contour in (\ref{e:invform}), has
been chosen to equal this unique positive solution.
Since the second derivative is $\overline{(H-\overline{H})^2} + \vep^2$,
it is now strictly positive, and the integration contour 
$\beta-\rmi\infty \to \beta+\rmi\infty$ 
in fact goes through the saddle point $\beta$ via the
path of steepest descent.  Using the saddle point approximation to
evaluate the contribution of this saddle point to the integral 
gives then
\[
 {1\over \sqrt{2\pi} }
 \biggl(\vep^2 + \overline{(H-\overline{H})^2}\biggr)^{-\half}
 \rme^{\half \beta^2 \vep^2 + \beta E} 
 \Tr\!\left(\ope{A}\rme^{-\beta\ham}\right).
\]
From this, equation (\ref{e:gid}) and the saddle point equation we
obtain the following exact result and its saddle point approximation
\begin{eqnarray}
 \Tr\!\left(\ope{A} \tihep(E)\right) &= 
 \rme^{-\half \beta^2\vep^2+\beta \overline{H}}
 \Tr\!\left(\ope{A} \tihep(\overline{H}) \rme^{-\beta\ham}\right)
    \label{e:spex} \\
 & \approx {1\over \sqrt{2\pi} }
 \biggl(\vep^2 + \overline{(H-\overline{H})^2}\biggr)^{-\half}
 \rme^{-\half \beta^2\vep^2+\beta \overline{H}}
 \Tr\!\left(\ope{A}\rme^{-\beta\ham}\right).\label{e:spapp}
\end{eqnarray}

So far we have only inspected the saddle points on the positive real
axis.  However, as a simple example with e.g.\ a harmonic oscillator
will show, there will generally be a countably infinite set of
saddle points on the complex plane.   Also, it is quite
possible that the steepest descent path going through all the relevant
saddle points will not stay on the right half-plane, which will be
unfortunate unless the analytical continuation 
of the canonical trace over the 
imaginary axis is known.  In addition, the values of $\beta$
at which the traces $\Tr\!\left(\ope{A} \tihep(E)\right)$ and
$\Tr\tihep(E)$ need to be evaluated are typically different, which would
then mean that the ratio of their saddle point approximations does not
exactly equal the canonical expectation value.

However, as we shall see in the next section, in the thermodynamical
limit these considerations are not relevant and the saddle point
approximation using the positive $\beta$ will give accurate results so
long as we use a suitable $\vep$ and inspect only large enough energies
$E$.  For the partition function this saddle point approximation reads
\begin{equation}\label{e:Zapp}
\ZG_{E,\vep}
 \approx {1\over \sqrt{2\pi (\sigma^2+\vep^2)}  }
 \rme^{-\half \beta^2\vep^2+\beta \overline{H}}  \Zcan_\beta,
\end{equation}
where $\overline{H} = \canmean{\ham}_\beta$, $\sigma^2 =
\canmean{\ham^2}_\beta - \overline{H}^2$ and $\beta$ is the unique
positive number satisfying $\canmean{\ham}_\beta = E + \vep^2 \beta$. 
Using this same $\beta$ also for the other trace in the expectation
values, will give the following exact result and an approximation of the
Gaussian expectation values,
\begin{eqnarray}\label{e:evsp}
 \gmean{\ope{A}}_{E,\vep} = 
 { \Tr\!\left(\ope{A} \tihep(\overline{H}) \rme^{-\beta\ham}\right)
  \over \Tr\!\left(\tihep(\overline{H}) \rme^{-\beta\ham}\right) }
 \approx \canmean{\ope{A}}_\beta.
\end{eqnarray}
From this formula we can already get some idea when the approximation
will be most accurate:  the contribution of the additional 
term $\tihep(\overline{H})$ will be small when the
variance of the Gaussian peak is greater than the variance of the
canonical distribution, i.e.\ whenever $\vep^2 > \sigma^2$.  
We will derive more quantitative bounds for the
accuracy of the approximation in the next section.

From the above result we can give a new interpretation of the
role of the canonical ensemble in quantum statistics: {\em canonical
ensemble is an approximation of the regularized
microcanonical ensemble from which
the discrete structure of the energy levels has been smoothened out.}\/ 
We have also seen that the canonical approximation works best in the
limit $\vep/\sigma\to\infty$.  However, for a fixed energy $E$ a minimum
requirement for the {\em Gaussian}\/ ensemble to give meaningful results
is $\vep < E -E_0$, so that taking this limit to its extreme 
is not possible in practice.  In
the next section we will propose a procedure for inspecting 
how and when a suitable compromise for $\vep$ can be found.

Finally, we would like to comment on the interpretation of the canonical
entropy, $\Scan_\beta = \beta \canmean{\ham}_\beta + \ln \Zcan_\beta$, 
in view of the previous approximation. Since the value of $\ZG_{E,\vep}$
gives the density of the energy eigenstates at energy $E$---density as
the number of eigenstates 
{\em per the energy interval $\vep$}---its logarithm
can be interpreted as the entropy of the system and we denote
$\SG_{E,\vep}=\ln \ZG_{E,\vep}$.
On the other hand, 
from the saddle point approximation (\ref{e:Zapp}) we conclude that
\begin{equation}\label{e:Sapp}
\SG_{E,\vep} \approx \Scan_\beta - \ln \vep 
 - \half\ln\left[2\pi\left(1+{\sigma^2\over\vep^2}\right)\right] 
 - \half \vep^2\beta^2,
\end{equation}
which means that the canonical entropy gives a good approximation
of the logarithmic
density of states per energy $\vep$ {\em provided}\/ the energy
resolution is proportional to the canonical energy deviation, i.e.\
$\vep\propto\sigma$.  If the energy resolution is microscopic, then the
above formula implies a correction to the canonical entropy
of the form $\ln N$, $N$ being the
number of particles.  Thus a natural
interpretation for the canonical entropy is the entropy measured from
the density of states per energy $\sigma$, the standard deviation
of the canonical ensemble---note that the relation between the density
of states and the dimensionless number given by the canonical
entropy is dubious in any case. 

\section{Efficiency and improvements of the canonical
ensemble\label{sec:imprcan}}

In this section we will continue working
with the same setup as in the
previous one, i.e.\ assume that the system is canonically bounded,
$\ope{A}$ is a canonical observable and the parameter $\beta$
is the unique positive solution to the equation 
$\canmean{\ham}_\beta = E + \vep^2 \beta$.
Since we will be mainly using
the canonical expectation values here, we will drop both the superscript 
``can'' and the subscript $\beta$ from these expressions;
similarly, using the ``hats'' to signify operators will become
cumbersome and we will abandon the practise at this point.

Our first aim is to derive quantitative bounds for how well the
canonical ensemble approximates the Gaussian one and then to derive a
method for computing corrections to the canonical ensemble when it first
begins to fail.  The corrections will be expressed in terms of the
normalized moments of the canonical distribution and for this reason we
will now adopt the notation $\sigma^2$ for the variance $\bigl\langle
\bigl(H-\mean{H}\bigr)^2\bigr\rangle$ and then define the normalized
Hamiltonian $h$ by the formula
\[
h = {H-\mean{H}\over \sigma}.
\]

If $A$ is positive and non-zero, then 
Jensen's inequality \cite{RCA} can be used for deriving the following 
bounds valid for any $a \in\R$,
\begin{equation}\label{e:Jensen1}
 \exp\biggl(-a {\mean{Ah^2}\over \mean{A}}\biggr) \le
 {\mean{A \exp(-ah^2)}\over \mean{A}} \le 1.
\end{equation}
If we apply this to 
$\mean{A\rme^{-ah^2}\!\exp(ah^2)}/\mean{A\rme^{-ah^2}}$,
 we can improve also the upper bound:
\begin{equation}\label{e:Jensen2} 
 \exp\biggl(-a {\mean{Ah^2}\over \mean{A}}\biggr) \le
 {\mean{A \exp(-ah^2)}\over \mean{A}} \le
 \exp\biggl[-a {\mean{Ah^2}\over \mean{A}}
 \exp\biggl(-a {\mean{Ah^4}\over \mean{Ah^2}}
  \biggr)\biggr].
\end{equation}
Since $\mean{h^2}=1$ by definition, we get for $A=\ope{1}$
\begin{equation}\label{e:Jensen3} 
  \exp(-a) \le \mean{\exp(-ah^2)} \le
 \exp\bigl[-a \exp\bigl(-a \mean{h^4}\bigr)\bigr].
\end{equation}

Applying these bounds with $a={\sigma^2\over 2\vep^2}$ to (\ref{e:evsp})
yields the following bounds for the relative efficiency of the canonical
expectation values
\begin{equation}\label{e:cevb1}
 \exp\biggl(
  {-{\sigma^2\over 2\vep^2} { \mean{Ah^2}\over \mean{A} } }
 \biggr) \le
  {\gmean{A}_{E_,\vep}\over \mean{A}} \le
 \exp\biggl( {\sigma^2\over 2\vep^2}\biggr),
\end{equation}
or by using the more accurate equation (\ref{e:Jensen2}),
\begin{eqnarray}\label{e:cevb2}
\fl
  -\ase \left[
    { \mean{Ah^2}\over \mean{A} } -
    \exp\left({-\ase\mean{h^4}}\right)
 \right] \le
  \ln {\gmean{A}_{E_,\vep}\over \mean{A}} \nonumber\\
 \le \ase \left[ 1 - 
    {\mean{Ah^2}\over \mean{A}} 
    \exp\left({ -\ase {\mean{Ah^4}\over \mean{Ah^2}} }\right)
 \right].
\end{eqnarray}
These equations prove, for positive observables $A$, the earlier alluded
statement that in the limit ${\vep/\sigma}\to \infty$ the approximation
by the canonical ensemble becomes exact.  On the other hand, the bounds
also suggest that it is necessary to have at least
${\vep^2\over\sigma^2} \lesssim \half$ before the canonical expectation
values give a trustworthy approximation of the Gaussian ones. 

For a fixed $E$ it is not in general possible to take the limit
$\vep\to\infty$, since then $\beta\to 0$ and thus also
$\sigma\to\infty$.  Besides,
it is also necessary to require that $\vep\ll E$ to get
a meaningful approximation to the microcanonical ensemble  
from the Gaussian one.  In practice, the most interesting applications
of the ongoing ideas are in the energy region in which the canonical
ensemble first begins to fail.  For this reason, we also need an
estimate {\em in terms of the canonical quantities}\/ for the region
where the canonical approximation is not reliable.

Since 
\begin{equation}\label{e:caneff}
E=\mean{H}-\beta\vep^2=\mean{H} \Bigl( 1 - 
{ \beta \sigma^2 \over \mean{H}} {\vep^2\over \sigma^2} \Bigr)
\end{equation}
should be positive,
we can now give the following rules for making an identification
between the canonical and the Gaussian ensemble
\begin{itemize}
\item The value of ${ \beta \sigma^2 \over \mean{H}}$ measures the
effectiveness of the canonical ensemble with a given $\beta$: the
smaller the value, the better the canonical ensemble works and
differences are expected to arise when it gets to be of the order of
one.  Note that in the thermodynamical limit, we have $\beta \propto
1/N$, $\mean{H}\propto N$ and $\sigma^2\propto N$ and therefore we would
expect this quantity to vanish as $1/N$, thus implying that the Gaussian
and the canonical ensemble become equivalent in the thermodynamical
limit. 
\item For those values of $\beta$ with ${\beta \sigma^2 \over \mean{H}}
\lesssim 1$, choosing an $\vep\propto\sigma$ will give the most reliable
results.  The proportionality factor need not be very large, since the
relative accuracy of the canonical approximation depends on the second
power of its inverse---for typical observables the accuracy can improve
even more quickly as can be seen from equation (\ref{e:cevb2}).  In any
case, we get from (\ref{e:caneff}) an absolute upper bound for the
possible values of the proportionality factor,
\[
{\vep\over\sigma} \le \sqrt{ \mean{H} \over \beta \sigma^2 }
\propto \sqrt{N}.
\]
\end{itemize}

Let us then derive an approximation for the Gaussian partition function
in the region where the parameter $a={\sigma^2\over 2\vep^2}$ is small. 
From (\ref{e:spex}) we get the following identity
\[
 \ZG_{E,\vep} =
 {1\over \sqrt{2\pi (\sigma^2+\vep^2)}  }
 \rme^{-\half \beta^2\vep^2+\beta \mean{H}} \Zcan_\beta
 \sqrt{1+2 a} \left\langle \rme^{-a h^2} \right\rangle,
\]
where we have extracted the saddle point approximation that was given in 
(\ref{e:Zapp}).
Let us denote the logarithm of the
correction to the saddle point approximation by 
$f(a) = \half\ln(1+ 2 a) + \ln \left\langle \rme^{-a h^2}\right\rangle$.
Equation (\ref{e:Jensen3}) immediately yields simple
bounds for this correction term,
\[
\half\ln(1+ 2 a) - a \le f(a) 
 \le \half\ln(1+ 2 a) - a \rme^{-a \mean{h^4}}.
\]

Clearly, $f$ is an analytic function on the right half plane and 
it is infinitely many times differentiable
from the right at the origin of the real axis.  Thus it has a
Taylor polynomial expansion at the origin,
\begin{equation}\label{e:ftaylor}
f(a) = \sum_{k=0}^{K-1} {a^k \over k!} f^{(k)}(0^+) + \order{a^K}
\text{, for all }a>0,
\end{equation}
although the corresponding full Taylor series 
{\em need not}\/ converge.  This situation is the same as is
often encountered in a perturbation theory: the result is an
asymptotic series in the perturbed coupling constant.

Since $f(0^+)=0$, the constant term of the expansion vanishes and, by
virtue of the saddle point approximation, also the first coefficient is
zero, since $\mean{h^2}=1$.  The rest of the coefficients of the
expansion (\ref{e:ftaylor}) are then given by the formula
\[
 f^{(k)}(0^+) = (-2)^{k-1} (k-1)! + 
   {\rmd^k\over \rmd a^k} \ln 
  \left.\left\langle \rme^{-a h^2}\right\rangle\right|_{a=0}
    \text{, for }k\ge 1,
\]
the first five of which are computed in table \ref{t:Zcoeff}.

\begin{table}
\caption{The first five derivatives needed in 
the Taylor expansion of the 
Gaussian partition function.  In the table, 
$h_n$ refers to the canonical expectation
value $\mean{h^n}$.\label{t:Zcoeff}}
\begin{indented}
\item[]\begin{tabular}{@{}ll}
\br
$k$ & $f^{(k)}(0^+)$ \\
\mr
1 & $0$ \\
2 & $-3 + h_4$ \\
3 & $6 + 3 h_4 - h_6$ \\
4 & $-54 + 12 h_4 - 3 h_4^2 - 4 h_6 + h_8$ \\
5 & $360 + 60 h_4 - 30 h_4^2 - 20 h_6  + 10 h_4 h_6
	+ 5 h_8 - h_{10}$ \\
\br
\end{tabular}
\end{indented}
\end{table}

\begin{table}
\caption{The first five general derivatives needed for
the Taylor residuals of the 
Gaussian partition function.  In the table, 
$g_n$ refers to the $\alpha$-dependent ratio
$\mean{h^n \exp(-\alpha h^2)}/
\mean{\exp(-\alpha h^2)}$.\label{t:RKcoeff}}
\begin{indented}
\item[]\begin{tabular}{@{}ll}
\br
$k$ & $f^{(k)}(\alpha)$ \\ 
\mr
1 & ${1\over 1 + 2\alpha} - g_2$ \\
2 & $-{2\over (1 + 2\alpha)^2} - g_2^2 + g_4$ \\
3 & ${8\over (1 + 2\alpha)^3} -  2 g_2^3 + 3 g_2 g_4 - g_6$ \\
4 & $-{48\over (1 + 2\alpha)^4} - 6 g_2^4 + 12 g_2^2 g_4
	- 3 g_4^2 - 4 g_2 g_6 + g_8$ \\
5 & ${384\over (1 + 2\alpha)^5} - 24 g_2^5 + 60 g_2^3 g_4
	- 30 g_2 g_4^2 - 20 g_2^2 g_6 + 10 g_4 g_6
	+ 5 g_2 g_8 - g_{10}$ \\
\br
\end{tabular}
\end{indented}
\end{table}

With a given $K$, the residual term $R_K(a)$---denoted by $\order{a^K}$
in (\ref{e:ftaylor})---can be written as 
\[
R_K(a) = {a^K \over K!} f^{(K)}(\alpha),
\]
where $\alpha\in [0,a]$ depends on $a$.  Since all derivatives are
continuous, the values of the derivatives at $0^+$ given in table
\ref{t:Zcoeff} give an estimate for 
$f^{(K)}(\alpha)$ for sufficiently small $a$. Therefore,
these values can be used
for estimating the residual term and thus for deciding the best
value of $K$ for a given, small, $a$.  

If $a$ is not small enough, it is possible to use other, less accurate,
estimates. Now $f^{(K)}(\alpha)$
is a polynomial of the ratios 
$g_n=\mean{h^n \exp(-\alpha h^2)}/\mean{\exp(-\alpha h^2)}$, which,
on the other hand, have the exact bounds
\[
\exp\Bigl( -a {h_{2 k +2}\over h_{2 k}}\Bigr) 
 \le { g_{2 k} \over h_{2 k} }
 \le \exp(a)\text{, for all }k \ge 1.
\]
Therefore, it is possible to use
these bounds in the known form of the
polynomials yielding exact bounds for the values of the derivatives.
The first five of the polynomials are given in table \ref{t:RKcoeff}
and the rest can be easily computed, if necessary.

The same Taylor polynomial approximation in $a$ can be made equally well
for expectation values of a canonical observable $A$.  Let us define
\[
g(a) = { \mean{A \exp(-a h^2)} \over \mean{\exp(-a h^2)} },
\]
when by (\ref{e:evsp}) the function $g(a)$ equals the Gaussian
expectation value of $A$ at energy $E=\mean{H}-\vep^2\beta$ and 
resolution $\vep=\sigma/\sqrt{2 a}$.  The Taylor expansion of $g$
is
\begin{equation}\label{e:gtaylor}
g(a) = \mean{A} \left( 1 + 
	\sum_{k=1}^{K-1} {a^k \over k!} G_k(0) 
	+ {\mean{A \exp(-\alpha h^2)}\over 
	\mean{A}\mean{\exp(-\alpha h^2)}}
	{a^K \over K!} G_K(\alpha) \right),
\end{equation}
where the coefficients $G_k$ are normalized derivatives,
\begin{equation}\label{e:Gkdef}
G_k(\alpha) = {\mean{\exp(-\alpha h^2)}\over
  \mean{A \exp(-\alpha h^2)} } g^{(k)}(\alpha),
\end{equation}
and we have given the first few of them in table \ref{t:ader}.
The expansion up to terms of order $a^2$ is therefore given by 
\[
\fl
g(a) = \mean{A} + \mean{(1- h^2) A} a + 
	\left[ \mean{(1- h^2) A} + \half \mean{
	  (h^4-\mean{h^4}) A} \right] a^2
	+ \order{a^3}.
\]

\begin{table}
\caption{The first few normalized derivatives---as defined in 
equation (\ref{e:Gkdef})---needed in 
the Taylor expansion of the Gaussian expectation values.  
Here $g_n$ and $A_n$ refer to the $\alpha$-dependent ratios
$\mean{h^n \exp(-\alpha h^2)}/\mean{\exp(-\alpha h^2)}$ and
$\mean{h^n A\exp(-\alpha h^2)}/\mean{A\exp(-\alpha h^2)}$,
respectively.\label{t:ader}}
\begin{indented}
\item[]\begin{tabular}{@{}ll}
\br
$k$ & $G_k(\alpha)$ \\ 
\mr
1 & $-A_2 +  g_2$ \\
2 & $A_4 - 2 A_2 g_2 + 2 g_2^2 - g_4$ \\
3 & $-A_6 + 3 A_4 g_2 - 6 A_2 g_2^2 + 6  g_2^3 + 3 A_2 g_4
	-6 g_2 g_4 + g_6$ \\
4 & $A_8 - 4 A_6 g_2 + 12 A_4 g_2^2 - 24 A_2 g_2^3 + 24 g_2^4
	- 6 A_4 g_4 + 24 A_2 g_2 g_4 - 36 g_2^2 g_4 
	+ 6 g_4^2$ \\
  & $\quad - 4 A_2 g_6 + 8 g_2 g_6 - g_8$ \\
5 & $-A_{10} + 5 A_8 g_2 - 20 A_6 g_2^2 + 60 A_4 g_2^3 -
	120 A_2 g_2^4 + 120 g_2^5 + 10 A_6 g_4 - 60 A_4 g_2 g_4$ \\
  & $\quad
	+ 180 A_2 g_2^2 g_4 - 240 g_2^3 g_4 - 30 A_2 g_4^2 
	+ 90 g_2 g_4^2 + 10 A_4 g_6 - 40 A_2 g_2 g_6 + 60 g_2^2 g_6$ \\ 
  & $\quad - 20 g_4 g_6 
	+ 5 A_2 g_8 - 10 g_2 g_8 + g_{10}$ \\
\br
\end{tabular}
\end{indented}
\end{table}

\section{Summary of the results}

In the first section we have shown that when $\vep\to 0$ the Gaussian
ensemble approaches the quantum microcanonical ensemble and the Gaussian
expectation values pick out the nearest microcanonical expectation
values.  This makes the Gaussian ensemble an easy to use approximation
of the microcanonical ensemble, although we have also argued that the
ensemble can be given an independent physical interpretation in certain
kinds of experiments. 

The main results of this paper, however, consider the opposite limit,
where $\vep$ is much larger than a typical distance between consecutive
energy levels.  We have shown how the canonical ensemble forms an
accurate approximation of the Gaussian ensemble in this limit.  On the
other hand, since the Gaussian ensemble is a regularization of the
discrete microcanonical ensemble in this limit, the canonical ensemble
can also be given an interpretation as an approximation of this
regularized microcanonical ensemble. 

If $E$ and $\vep$ are the Gaussian energy and energy resolution,
respectively, then there always exists a unique $\beta>0$ defined by the
equation $\canmean{H}_\beta = E + \beta\vep^2$ and this $\beta$ gives
the best inverse temperature for a canonical approximation of the
Gaussian ensemble.  This approximation is best characterized by the
parameter $a=\sigma^2/2\vep^2$, where $\sigma^2$ is the variance of the
energy in the canonical ensemble, $\sigma^2 = \canmean{H^2}_\beta -
(\canmean{H}_\beta)^2$.  The approximation was shown to work at least in
the region where $a\ll 1$ and, since it was necessary to have
$a\gtrsim\beta\sigma^2/\canmean{H}_\beta$, the latter quantity furnishes
an indicator of whether or not the canonical ensemble can be used for
getting reliable information about the properties of a closed system. 
This indicator can also be given in the form
$\beta\sigma^2/\canmean{H}_\beta = C_V/\beta\canmean{H}_\beta$, where
$C_V$ is the specific heat at constant volume. 

Quantitatively, we have proven the following formulas for the canonical
approximation of the Gaussian ensemble
\begin{eqnarray}
\SG_{E,\vep} = \Scan_\beta - \ln\vep
 -\half\ln\left(2\pi\right) - {\sigma^2\beta^2 \over 4 a} 
 -\half\ln\left( 1+ 2 a\right) 
 + \Delta S(a;\beta), \\
\gmean{A}_{E,\vep} = \canmean{A}_\beta + \Delta A(a;\beta).
\end{eqnarray}
The correction terms, which become important in the region $a\approx 1$,
can be given an asymptotic expansion in terms of the 
moments of the normalized canonical energy operator, 
$h = (H-\mean{H})/\sigma$,
\begin{eqnarray}
\fl \Delta S(a;\beta) = \half (\mean{h^4} - 3) a^2
	+ {1\over 6} (6 + 3 \mean{h^4} - \mean{h^6}) a^3 
	+ \order{a^4}, \\
\fl \Delta A(a;\beta) = \mean{(1- h^2) A} a + 
	\left[ \mean{(1- h^2) A} + \half \mean{
	  (h^4-\mean{h^4}) A} \right] a^2
	+ \order{a^3}.
\end{eqnarray}
The entropy deviation has also the exact bounds
\begin{equation}
\half\ln(1+ 2 a) - a \le \Delta S(a;\beta)
 \le \half\ln(1+ 2 a) - a \exp({-a \mean{h^4}}),
\end{equation}
while for {\em positive}\/ observables $A$ we have derived bounds 
for the logarithmic proportional deviation
\begin{eqnarray}
\fl
  -a \left[ { \mean{Ah^2}\over \mean{A} } -
    \exp\!\left({-a\mean{h^4}}\right) \right] 
 \le \ln {\gmean{A}_{E_,\vep}\over \canmean{A}_\beta } \le
 a \left[ 1 - {\mean{Ah^2}\over \mean{A}} 
    \exp\!\left({ -a {\mean{Ah^4}\over \mean{Ah^2}} }\right)
 \right].
\end{eqnarray}
All expectation values in the above are in the canonical ensemble
unless stated otherwise.

\section{Thermodynamics of systems with exponentially
increasing density of states}

We will now repeat the analysis done in sections \ref{sec:cangauss} and
\ref{sec:imprcan} on a system with exponentially increasing density of
states.  In this case, the canonical ensemble can be defined only up to
a certain value of the inverse temperature $\beta$ and it is not clear
when and if the canonical ensemble will give meaningful results. 
However, since there are physically interesting systems which exhibit an
exponential increase of the density of states, e.g.\ free bosonic string
theory \cite{GSW1}, and since computations in the canonical ensemble are
typically easier to perform than microcanonical ones, it is useful to
know if the canonical ensemble can be applied to analysis of such a
system. 

We first define what is meant by an exponential increase of the density
of states: if there exists a finite number
\[
\beta_c = \inf\defset{\beta>0}{\tr \rme^{-\beta H} < \infty},
\]
then $\beta_c$ can be identified with the speed of exponential increase
of the density of states as clearly $\tr \rme^{-\beta H} < \infty$ for
all $\beta>\beta_c$ and $\tr \rme^{-\beta H} = \infty$ for
$\beta<\beta_c$.  Also, in the following we will consider only
observables $A$ with the property $\tr |A| \rme^{-\beta H} < \infty$ for
all $\beta>\beta_c$.  Since we have assumed that $\beta_c<\infty$, the
system is Gaussianly bounded and $A$ is a Gaussian observable.  Note
also that $\beta_c = 0$ corresponds precisely to the canonical case. 

Under these assumptions, everything said in the beginning of section
\ref{sec:cangauss}, especially formulas
(\ref{e:gid})--(\ref{e:invform}), still hold if we only require
$\beta>\beta_c$ instead of $\beta>0$.  Similarly, the analyticity of the
integrand in (\ref{e:invform}) is guaranteed only in the half plane $\re
w > \beta_c$.  In the saddle point approximation of this integral, the
uniqueness of the positive saddle point still holds with the same proof
as before, but the {\em existence} depends crucially on the behaviour of
the canonical ensemble at temperature $1/\beta_c$ or, more specifically,
on the value of $E_c \equiv \lim_{\beta\to\beta_c^+} \canmean{H}_\beta$. 

If $E_c=\infty$, then there exists
for all $E\in\R$ and $\vep>0$ a unique $\beta>\beta_c$ for which 
$\canmean{\ham}_\beta = E + \beta \vep^2$. If $E_c < \infty$, then there
is a positive saddle point $\beta$ if and
only if $E<E_c-\beta_c\vep^2$.  In these two cases everything said in
section \ref{sec:imprcan} will hold for the saddle point value of
$\beta$ and thus it is reasonable to use canonical ensemble for those
systems with $C_V/\canmean{\beta H}_\beta\ll 1$.

The situation is different for those values of $E$, for which 
$E\ge E_{\rm max}\equiv E_c-\beta_c\vep^2$.  
Then there are no saddle points on the
positive real axis and the best one can do with the canonical
ensemble is to choose $\beta\to\beta_c$.  This approximation, however,
is not good unless $E\approx E_{\rm max}$
as can be seen from the relation
\[
 \gmean{A}_{E,\vep} = 
 { \Tr\!\left(A \rho_\vep(E+\beta_c\vep^2) \rme^{-\beta_c H}\right)
  \over \Tr\!\left(\rho_\vep(E+\beta_c\vep^2) 
  \rme^{-\beta_c H}\right) }
 = { \canmean{A \rho_\vep(E_c+E-E_{\rm max})}_{\beta_c} \over
	\canmean{\rho_\vep(E_c+E-E_{\rm max})}_{\beta_c} },
\]
which is a consequence of equation (\ref{e:gid}).

Thus we have found that if $\canmean{H}_{\beta_c}=\infty$ the system can
always be approximated by the canonical ensemble in the thermodynamical
limit just as has been explained in the preceding sections.  In effect,
as the system is ``heated up'' by adding more and more energy to it, the
temperature of the system will increase asymptotically to the limiting
value $T_c=1/\beta_c$.  On the other hand, if
$\canmean{H}_{\beta_c}<\infty$, then there exists a limiting energy
$E_{\rm max}\approx\canmean{H}_{\beta_c}$ after which the canonical
ensemble should not be used for estimating the statistical behaviour of
the system but using some form of the microcanonical ensemble instead
would be advisable. 

\section{Discussion}

In this paper we have examined relations between the microcanonical and
the canonical approach to quantum statistics.  On the level of ideas,
the present results are well-known and well-presented in most standard
textbooks on statistical physics and, naturally, none of our results
tells anything new about systems fully in the thermodynamical limit. 
What we have aspired to do here is to develop a systematic treatment of
systems, which are neither large enough to be considered thermodynamical
nor simple enough to be completely solvable, but which are,
nevertheless, in an energetic equilibrium with their environment. 

By taking the energy fluctuations of the system as an integral part of
the microcanonical formalism we have been able to show rigorously how
the canonical ensemble gives an approximation of the microcanonical
statistics even for systems which are not even near a thermodynamical
limit and we have been able to give precise relations between canonical
concepts, such as temperature, and the more fundamental concepts related
directly to energy.  We have also shown how and when the canonical
ensemble can be stretched to aid in the analysis of these non-thermal
systems. 

In our analysis of regularization of the energy spectrum we have limited
ourselves to the Gaussian distribution.  This, however, is only a
convenient choice and the analysis could be repeated by using any smooth
function with a compact support instead.  Using this second alternative
would, in fact, be necessary for microcanonically bounded systems with
the logarithm of the density of states increasing faster than
quadratically in energy, but we will not redo this analysis here. 
Neither have we yet discussed the evaluation of the canonical moments of
the energy operator, which are required for the asymptotic expansion of
the Gaussian expectation values---this will be the subject of a
subsequent work, where we will also show how lattice Monte Carlo methods
can be employed in evaluation of the Gaussian expectation values. 

For sake of mathematical definiteness, we have in this paper inspected
only quantum mechanical systems, but on the level of formal
manipulation, the results given here can be equally well interpreted as
statements about statistical quantum {\em field}\/ theory.  Since
canonical quantum field theory is already a well-developed part of
physicists' toolkit \cite{bellac:tft}, this is not as bold a claim as it
looks at first sight.  In fact, the formulation of quantum
microcanonical ensemble given in \cite{chs:93} can also be obtained from
the $\vep,\beta\to 0^+$ limit of the Gaussian formula (\ref{e:invform}),
which might then lead to a more rigorous derivation of that formulation
after the intricacies in the definition of four-dimensional statistical
quantum field theories have been resolved. 

\ack

I would like to thank Masud Chaichian, Antti Kupiainen, Claus Montonen
and Esko Suhonen for patient commentary and reading of the manuscript
and Sami Virtanen and Arttu Rajantie for several useful discussions.

\end{document}